\documentclass[aps,pra,twocolumn,superscriptaddress,showpacs]{revtex4-1}

\usepackage{graphicx}
\usepackage{dcolumn}
\usepackage{bm}

\begin{document}

\title{Multiple Rabi Splittings under Ultra-Strong Vibrational Coupling}
\author{Jino George}
\affiliation{ISIS \& icFRC, Universit\'e de Strasbourg and CNRS, Strasbourg, France}

\author{Thibault Chervy}
\affiliation{ISIS \& icFRC, Universit\'e de Strasbourg and CNRS, Strasbourg, France}

\author{Atef Shalabney}
\affiliation{ISIS \& icFRC, Universit\'e de Strasbourg and CNRS, Strasbourg, France}
\affiliation{Braude College, Snunit St 51, Karmiel, 2161002 Israel}
\author{Elo\"ise Devaux}
\affiliation{ISIS \& icFRC, Universit\'e de Strasbourg and CNRS, Strasbourg, France}

\author{Hidefumi Hiura}
\affiliation{Smart Energy Research Laboratories, NEC Corporation 34 Miyukigaoka, Tsukuba 305-8501, Japan}

\author{Cyriaque Genet}
\email[Electronic address: ]{genet@unistra.fr}
\affiliation{ISIS \& icFRC, Universit\'e de Strasbourg and CNRS, Strasbourg, France}

\author{Thomas W. Ebbesen}
\email[Electronic address: ]{ebbesen@unistra.fr}
\affiliation{ISIS \& icFRC, Universit\'e de Strasbourg and CNRS, Strasbourg, France}

\date{\today}

\begin{abstract}
From the high vibrational dipolar strength offered by molecular
liquids, we demonstrate that a molecular vibration can be
ultra-strongly coupled to multiple IR cavity modes, with Rabi
splittings reaching $24\%$ of the vibration frequencies. As a proof of
the ultra-strong coupling regime, our experimental data unambiguously
reveal the contributions to the polaritonic dynamics coming from the 
anti-resonant terms in the interaction energy and from the dipolar
self-energy of the molecular vibrations themselves. In particular, we
measure the opening of a genuine vibrational polaritonic bandgap of
ca. $60$ meV. We also demonstrate that the multimode splitting effect
defines a whole vibrational ladder of heavy polaritonic states
perfectly resolved.  
These findings reveal the broad possibilities in the vibrational ultra-strong
coupling regime which impact both the optical and the molecular 
properties of such coupled systems, in particular in the context of
mode-selective chemistry. 
\end{abstract}


\maketitle

Light-matter interactions in the strong coupling regime offer exciting 
possibilities for exploring quantum coherent effects both from a physical 
and chemical perspectives. This regime can be reached when a confined 
electromagnetic field interacts coherently with an electronic transition 
of an embedded material, leading to the formation of polaritonic states \cite{Agranovich}. Many realizations of this effect have 
been demonstrated, ranging from single atoms \cite{HarocheReview},
quantum wells \cite{Skolnick, Bloch}, superconducting q-bits \cite{Wallraff}
to molecular systems \cite{Pockrand1982, Lidzey,Holmes,Bellessa,Dintinger,MahrtBEC, JinoConductivity,TormaBarnes}. Recently, we demonstrated that
molecular vibrations can be 
strongly coupled to an optical mode of a
Fabry-P\'erot (FP) cavity in the infrared (IR) region
\cite{ShalabneyNature,JinoLiquidPhase}. 
Such coupling is attracting more and more attention \cite{Tischler,
  Simpkins, Spanish} since molecular
vibrations play a key role in chemistry. Therefore, vibrational strong
coupling could potentially be used to control chemical reactions in
the same way as it has been demonstrated for electronic strong
coupling \cite{JamesChemReaction}. Lately, a whole field of research has been opened with the
prediction and demonstration of an ultra-strong coupling (USC)
regime \cite{Ciuti}. 
The USC regime indeed leads to
the possibility of probing fascinating properties of the coupled states
such as non-classical ground state behavior, squeezed vacuum and
polaritonic bandgaps \cite{Todorov, Faist, Ashkenzi, Gambino, Tal}. 

In this Letter, we demonstrate that USC can also be reached, at room
temperature, with ground state molecular vibrations coupled to an
optical mode. Inherent features of the USC regime, anti-resonant and
self-energy contributions to the coupling, are measured on vibrational
polaritonic states. These results reveal totally different dynamics than the one
we previously reported for vibrational strong
coupling \cite{ShalabneyNature}. To reach this vibrational USC regime, we
exploit unique features of molecular liquids with high vibrational dipolar strength.
Such liquids ressemble assemblies
of individual ground-state mechanical oscillators where one is able to
reach dipolar strength densities far beyond the quenching densities in the solid
phase. These combined features naturally set the conditions for
collective coupling strengths up to the USC
regime. Remarkably, with Rabi
splitting practically matching the free spectral range
(FSR) of the IR cavity, the coupling process
involves multiple orders of the FP cavity modes and leads to a genuine ladder of
polaritonic states. We show multimode splitting with up to 10 polaritonic peaks.
The new vibrational spectrum associated
 with this polaritonic ladder differs radically from the vibrational
 spectrum of the bare molecules. 
While USC has been recognized as a new playground for electronic
polaritonic physics, the USC features that we now observe on vibrations
are expected to also have a strong impact on the chemistry of vibrational
polaritonic states that remains so far unexplored.

\begin{figure*}
\includegraphics[width=1.9\columnwidth]{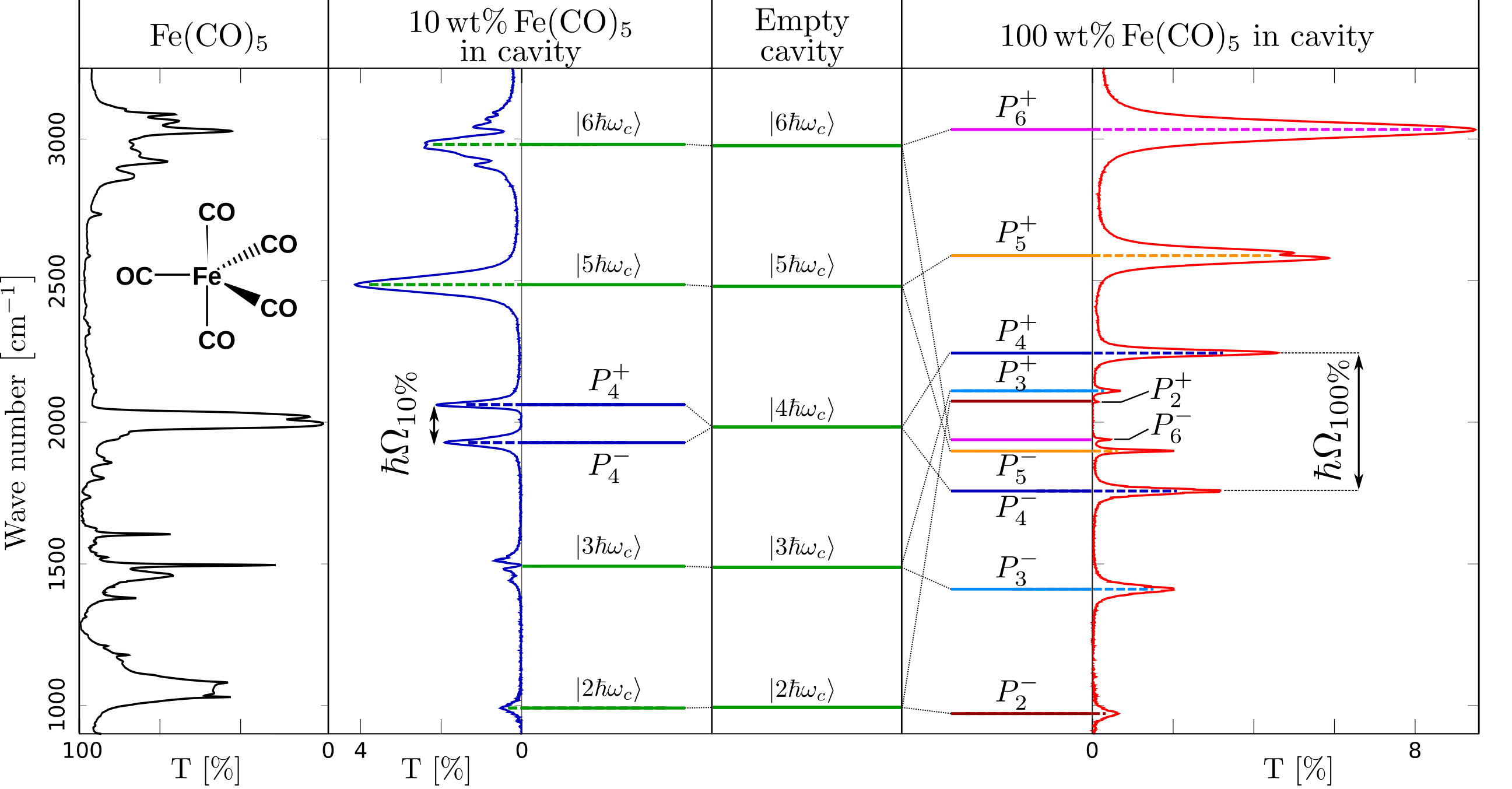}
\caption{Fe(CO)$_5$ data are shown in columns for clarity,
  starting with the IR spectrum of 10 wt \% Fe(CO)$_5$ in toluene,  the
  transmission spectrum of the FP cavity filled with the same solution
  (blue curve) with a Rabi splitting
  $\Omega_{10\%}\sim\!135\,\textrm{cm}^{-1}$, followed by the mode
  diagram of the system under the corresponding coupling.  The middle
  column shows the empty cavity modes with, on the
  right-hand side, the coupled diagram of the
  multiple polaritonic states when the cavity is filled with 100 wt \% Fe(CO)$_5$.
  The last column to the right shows the corresponding experimental IR spectrum of
  the filled cavity with a resonant Rabi splitting
  $\Omega_{100\%}\sim\!480\,\textrm{cm}^{-1}$. \label{fig:fig1}}. 
\end{figure*}

Our system consists of a micro-fluidic FP flow cell 
which can be filled with any given molecular liquid (see Appendix \ref{AA}). It is made of two 
ZnSe windows coated with $13\,\textrm{nm}$ thick Au films to form the FP mirrors and 
closing them with a Mylar spacer of the appropriate thickness produces a microcavity 
with IR modes of quality factors $Q\sim\!50$ \cite{JinoLiquidPhase}. 
By varying the spacer thickness, one of the 
optical modes is brought into resonance with the targeted molecular
vibration. Different concentrations of molecules are injected 
in the cell and the system is spectroscopically characterized using 
a commercial Fourier Transform IR spectrophotometer (FTIR, Nicolet-6700).  
The two molecules chosen for this study are the iron pentacarbonyl 
Fe(CO)$_5$ (see Fig.~\ref{fig:fig1}) and carbon disulphide CS$_2$. Fe(CO)$_5$ liquid has a very strong oscillator
strength with 3 equitorial and 2 axial CO-stretching degenerate
modes having a fundamental frequency $\omega_\nu$ corresponding to a
wave number of $\sim\!2000\,\textrm{cm}^{-1}$ \cite{Cataliotti}. The IR absorption
band of a dilute Fe(CO)$_5$ solution (10 wt \% in toluene) is shown in
Fig.~\ref{fig:fig1}. We inject the same solution into a 
FP cavity tuned to have its $4^{th}$-order longitudinal mode resonant with the CO-stretching band. This resonant coupling 
splits the fundamental vibrational mode into an 
upper and lower mode separated by $\hbar\Omega_{10\%}\sim\!135\,\textrm{cm}^{-1}$
(Fig.~\ref{fig:fig1}, second column). Importantly, this mode splitting is 
larger than both the width of the cavity mode and of the vibrational
peak, i.e. it corresponds to a genuine Rabi splitting.

Using pure Fe(CO)$_5$ liquid under the same resonant conditions expectedly leads to an increase in the mode splitting up to $\hbar\Omega_{100\%}\sim\!480\,\textrm{cm}^{-1}$, as shown in Fig.~\ref{fig:fig1} (fourth column). In these conditions, the vibrational spectrum of the
coupled Fe(CO)$_5$ liquid also displays a series of sharp
resonances that stem from the coupling of the CO-stretching band with successive longitudinal modes of the FP cavity.
We report in Appendix \ref{AB} similar spectral
evolutions for CS$_2$. This multimode splitting theoretically predicted
 by Meiser and Meystre \cite{Meystre} is similar to
that reported in the case of electronic strong coupling \cite{Dupont,
  Yu}.

In order to understand the multi-peaked structure of the spectrum, we perform a transfer matrix simulation on a cavity filled
with pure Fe(CO)$_5$ liquid. Solving
the multilayered structure consisting of the
ZnSe flow cell windows, the Au cavity mirrors and the embedded
absorbing medium, the calculated cavity transmission spectrum is shown together
with the measured spectrum in Fig.~\ref{fig:fig2}(a). A detailed
description of the modeling of our system is given in Appendix \ref{AA}. The electric field distribution inside the cavity was
computed using the same parameters and is shown in Fig.~\ref{fig:fig2}(b). As can
be seen from the field distributions, the CO-stretching mode of Fe(CO)$_5$, when resonantly coupled to the $4^{th}$-order mode of the FP cavity, gives rise to an upper and a lower mode, at $2245\,\textrm{cm}^{-1}$ and $1756\,\textrm{cm}^{-1}$
respectively. The other four new resonances on either sides of the
fundamental CO-stretching mode are at $2110, 2071, 1938, 1898\,\textrm{cm}^{-1}$ (the
other peaks outside this spectral window can be seen in Fig.~\ref{fig:fig1}). The field distributions enable us to identify the modes at higher
energy as originating from the coupling between the vibrational band
and lower optical modes of the cavity and vice versa for the modes at lower
energy.

The positions of these modes is directly given by computing the round trip phase accumulation $\delta\phi=2L\omega n/c + 2\phi_r$ for the
electromagnetic field in the cavity, where $L$ is the
cavity length, $\omega$ is the vacuum frequency of light, $c$ is the
speed of light, $n$ is real part of the refractive index and $\phi_r$
is the reflection phase due to the finite metal skin depth
\cite{Zhu}. As shown in Fig.~\ref{fig:fig2}(c), the dispersive character of the pure molecular liquid is so strong that optical modes of different $m^{th}$-orders can satisfy simultaneously the resonant phase condition $\delta\phi=2\pi m$, with two solutions $P^+_m, P^-_m$ for each mode $m$. 
The observed vibrational ladder
is characterized by very large multi-mode splittings, with $\Omega_{100\%}$ reaching ca. $24\%$
of the vibrational mode energy. Such a high ratio is often encountered in ultra-strongly coupled systems, and in order to confirm that our molecular liquid has entered into the USC regime, we now show that the spectral structure of the coupled vibrational ladder cannot be described outside the framework of ultra-strong light-matter interaction.

\begin{figure}
\includegraphics[width=0.9\columnwidth]{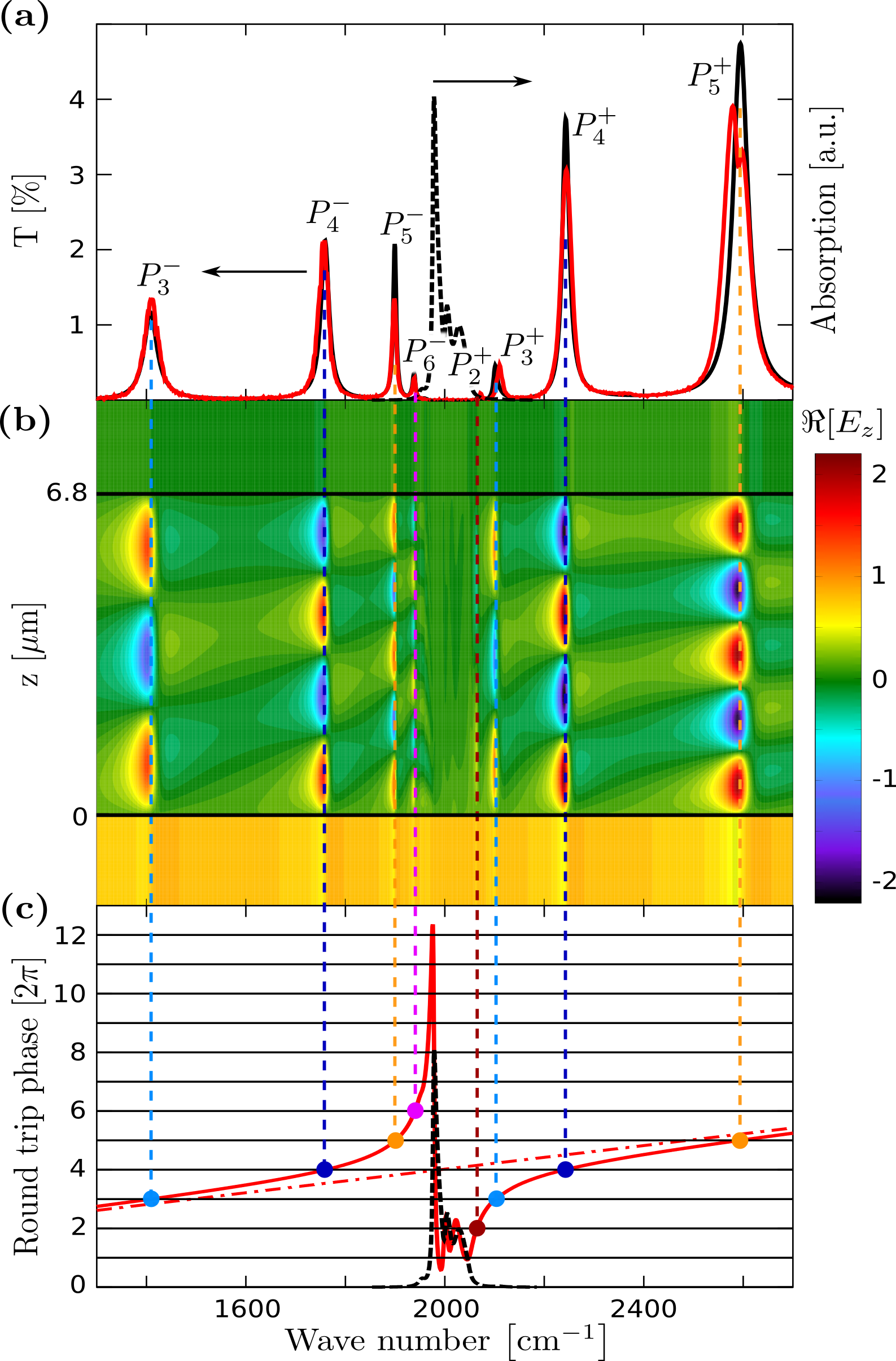}
\caption{(a) Experimental (red line) and transfer matrix simulated
  (black line) IR transmission spectra of pure Fe(CO)$_5$ coupled to
  the $4^{th}$-order mode of a flow cell FP cavity; black dashed line is the
  simulated absorbance of CO stretching mode of pure Fe(CO)$_5$. (b)
  T-matrix simulation of the electric field distribution inside the FP
  cavity. Asymmetric field distribution on either side of the
  vibrational band is an indication of higher and lower mode folding
  effects due to strongly dispersive refractive index. (c) Round trip
  phase accumulation of the electromagnetic field in the cavity with
  (red line) and without (red dashed line) absorber. Optical
  resonances occur for phase accumulation equal to integer multiples
  of $2\pi$ (horizontal black lines). The dispersion of the refractive
  index of the intra-cavity medium allows multiple solutions for
  various mode indices (vertical dashed lines, same color code as in
  Fig.~\ref{fig:fig1}). The fitted
  absorption line shape of Fe(CO)$_5$ is also shown in black dashed
  line. The resonances lying in the strong absorption region are over
  damped solutions and do not appear in the transmission spectra. \label{fig:fig2}}
\end{figure}

\begin{figure}
\includegraphics[width=\columnwidth]{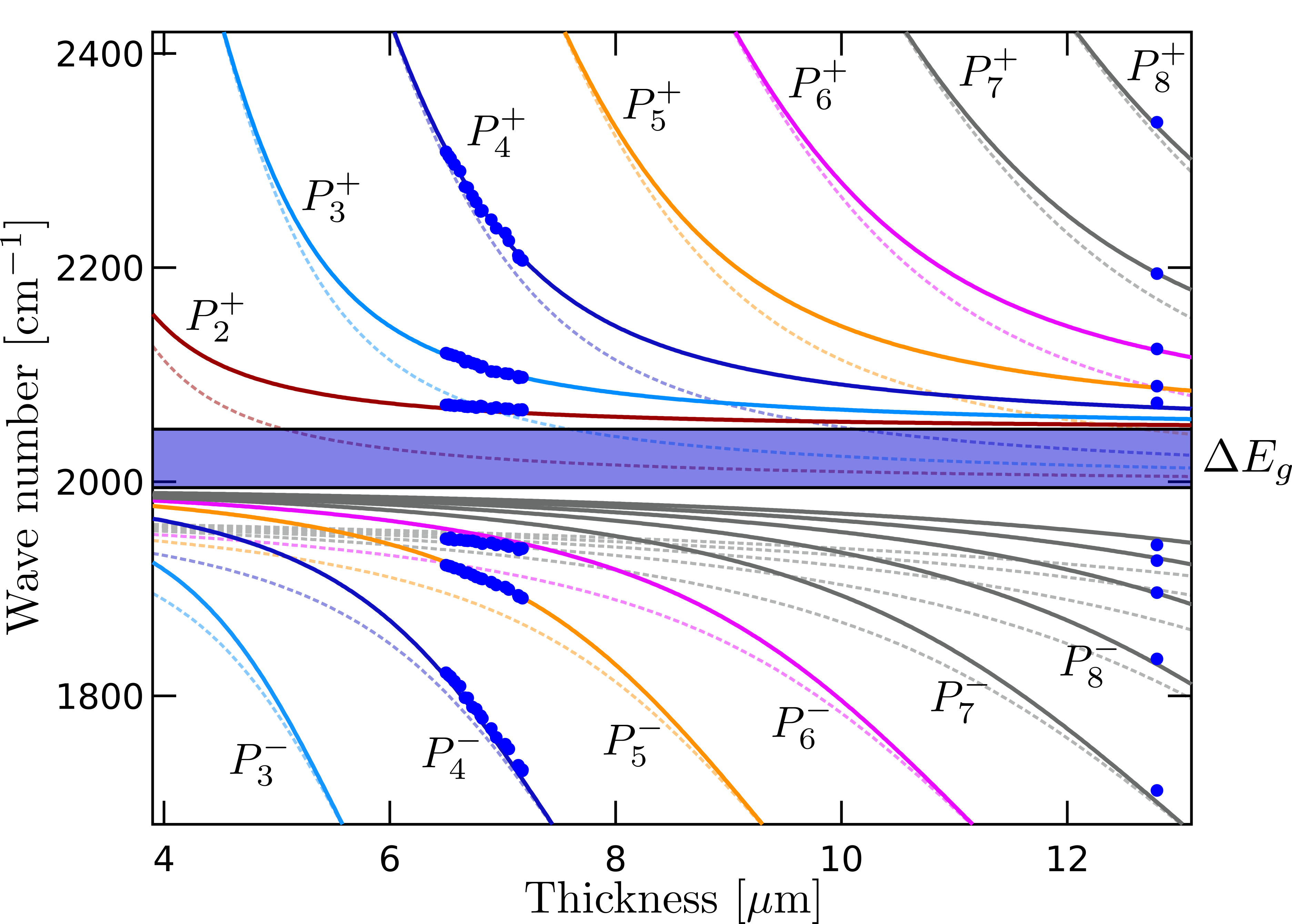}
\caption{Polaritonic dispersion diagram as a function of the cavity
  thickness. Experimental transmission peaks positions are reported as
  blue dots. The exact cavity thickness is determined from the free
  spectral range of the cavity -see Appendix \ref{AA}. The solid lines are the best
  fit to the data using the full Hamiltonian model that accounts for the dipolar self-energy and the contributions beyond the rotating wave approximation (RWA), while
  the dashed lines are solutions of the simple RWA model -see Appendix \ref{AD} for the fitting procedure. The color code is the
  same as in Fig.~\ref{fig:fig1}. The vibrational polaritonic
  bandgap $\Delta E_g$ only appears in the full Hamiltonian model.\label{fig:fig3}}
\end{figure}

The involvement of specific features of the USC regime can be revealed most directly at the level of polaritonic dispersion diagrams. We have measured the dispersions of our coupled
vibrational modes as a function of the cavity
thickness (i.e. as a function of the detuning). This is done
by varying the cavity thickness around a fixed value determined
by the spacer inserted in our flow cell. The results are gathered in
Fig.~\ref{fig:fig3}. Using a thicker spacer, we obtain the asymptotic
positions of the coupled modes. These experimental data are compared
to polaritonic dispersions that are calculated from a coupled oscillator model that takes explicitly into account the contributions
of the vibrational dipolar self-energy in the molecular liquid and the
anti-resonant coupling terms which are specific to the USC regime (see Appendix \ref{AC} for a detailed presentation of the model). We emphasize that in our model, there is only one free parameter which corresponds to the
Rabi splitting $\hbar\Omega_R$. 

As discussed in Appendix \ref{AD}, a coupled
oscillator model keeping only the resonant interaction terms at 
$\mathcal{O}(\Omega_R/\omega_\nu)$ order, and therefore neglecting the dipolar self-energy of the vibration (Jaynes-Cumming-type Hamiltonian), is unable to
fit accurately the experiment close to the bare vibrational mode
  energy, as shown in Fig.~\ref{fig:fig3} (blow up) and in full scale in the SM. This mismatch proves that our system is truly in  
the USC regime. Remarkably, the asymptotic values of the model
yield a vibrational polaritonic bandgap of $\sim60~\textrm{cm}^{-1}$,
as shown in Fig.~\ref{fig:fig3}. The opening of such a bandgap is an
indisputable signature of the USC regime, as pointed out in the
context of intersubband electronic transition systems
\cite{Gambino,Ashkenzi}. Here, we emphasize again that such a
signature is observed for molecular vibrational transitions.  
More interestingly, the USC regime also implies that the vibrational ground
state must shift to lower energies 
while acquiring a photonic admixture.
Such modifications are analogous to those described at the level of
electronic transitions, except that in our case, they
imply a deep modification of the whole vibronic landscape of the
dressed molecules.

\begin{figure}
\includegraphics[width=\columnwidth]{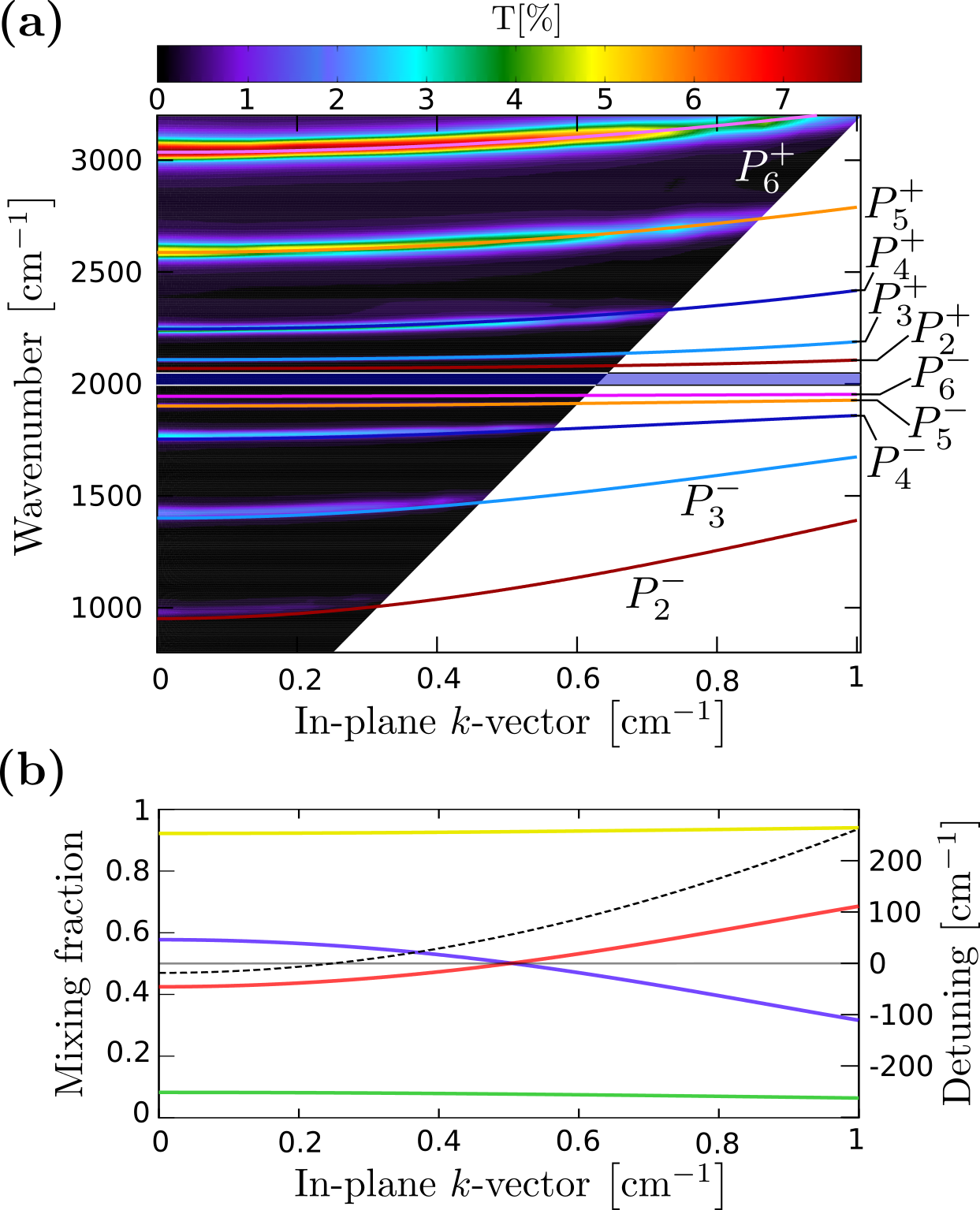}
\caption{(a) Polaritonic dispersion diagram measured by angle dependent IR transmission
spectroscopy (unpolarized, $0-28^\circ$). The solid lines
are the solutions of the full Hamiltonian model, using the same
parameters and color code as in Fig.~\ref{fig:fig3}. The vibrational
polaritonic bandgap is again clearly observed (blue horizontal band). (b) Photonic and
vibrational fractions of the $4^{th}$ polaritonic branch (blue and
red curves respectively) and of the $6^{th}$ polaritonic branch (green and
yellow curves respectively). The photon-vibration energy detuning for
the $4^{th}$ cavity mode is shown in black dashes (right axis). 
  \label{fig:fig4}} 
\end{figure}

The second remarkable feature of the vibrational dressed states under USC is related to the presence of a genuine ladder of vibrational polaritonic states.
To get further insight into the nature of these multiple polaritonic
states, we perform angle-dependent experiments. The dispersion diagram shown in Fig.~\ref{fig:fig4}(a) clearly demonstrates the dispersive
behavior of the different polaritonic states. This behavior
is inherited from the photonic component of the polaritonic states, and is the
signature of their hybrid light-matter nature. Taking the Rabi
frequency parameter extracted from the best fit of the polaritonic
thickness-dependent dispersions of Fig.~\ref{fig:fig3}, our USC
oscillator model perfectly matches these experimental angular
dispersion data. Again, the polaritonic bandgap is clearly
seen, demonstrating that this forbidden energy band exists for
any cavity thickness and at any angle.
 We stress that our model assumes no interaction between the
 polaritonic branches associated 
with different (orthogonal) cavity modes. However, non-trivial cross-talk between the different polaritonic branches should be expected when
accounting for the non-Markovian behavior of our system
($\Omega_{100\%}=2k_{\rm B}T$ \cite{Antoine}). This analysis however goes beyond the scope of this work.
The results of our fit are shown in Fig.~\ref{fig:fig4}(a), and the
extracted Hopfield coefficients for the polaritonic
states $P^-_4$ and $P^-_6$ are shown in Fig.~\ref{fig:fig4}(b) \cite{Yamamoto}. As
expected, the vibrational content
of the states increases as they approach the energy of the bare
vibrational mode. It is interesting to note the non-trivial evolution
of the Hopfield coefficients calculated for the lower $P^-_4$
polaritonic states which becomes more photon-like at resonance. This
unbalanced matter- vs. photon-like mixing fraction is another
remarkable feature of the USC regime that comes in clear contrast with
the usual regime of strong coupling. Between $P^+_4$ and $P^-_4$, the
ladder consists of heavy (i.e. large vibrational content) polaritonic
states. Surprisingly, those heavy polaritonic states display 
linewidths up to 5 times narrower than the width of the bare cavity mode,
and up to 6 times smaller than the linewidth of the bare (inhomogeneoulsy 
broadened) molecular vibration.  Because of the opening of the vibrational 
polaritonic bandgap, these heavy polaritonic states are pushed away from 
the dissipative region of the bare vibration, therefore remaining perfectly 
resolved with their narrow linewidths. 
The concomitance of multi-mode and ultra-strong coupling of 
vibrational modes can hence be seen as an interesting way to overcome
a major hurdle encountered in the physics of electronic strong
coupling \cite{Richard}. 

In summary, we have demonstrated that it is possible to reach the regime of USC in the vibrational realm. This is done using high oscillator
strength molecular liquids. We have revealed indisputable signatures of the USC regime, showing how the features inherent to the USC regime can be also found at the level of molecular vibrational modes.
Remarkably, the molecular polaritonic multi-mode folding shown here
is a practical way for generating heavy polaritonic states with smaller linewidths than both the optical transition and the molecular vibration, leading to enhanced coherence time. 
Perhaps more importantly, these results point to the potential impact
of the USC regime in the context of bond-selective chemistry. As we
already proposed \cite{JamesChemReaction, ShalabneyNature}, the dynamics of 
bond breaking in the ground state could be significantly 
modified by vibrational strong coupling and even more under the USC
regime where the whole vibrational ladder dressed by the IR cavity
field is redefined. All these features will no doubt impact both the optical, the molecular and the material
properties of these ultra-strongly coupled systems \cite{EspagnolsPRX}, 
enriching the possibilities offered by such light-matter interactions.

\begin{acknowledgments}
We thank David Hagenmuller for fruitfull discussions. This work was supported in part by the ERC Plasmonics no 22577, the ANR Equipex Union (ANR-10-EQPX-52-01), the Labex NIE projects (ANR-11-LABX-0058-NIE) and USIAS within the Investissement d'Avenir program ANR-10-IDEX-0002-02. 
\end{acknowledgments}

{\it Author Contributions -}
J.G. and T.C. contributed equally to this work.

\appendix

\section{Transfer matrix simulations}  \label{AA}

Transfer matrix simulations amounts to solving the classical problem
of a multi-layered stack of dispersive media in terms of forward and backward
propagating electric field amplitudes. Thus for a given system to
simulate, one needs to know the (complex) refractive indices and the
thicknesses of each of the layers. 

\begin{figure}
\includegraphics[width=\columnwidth]{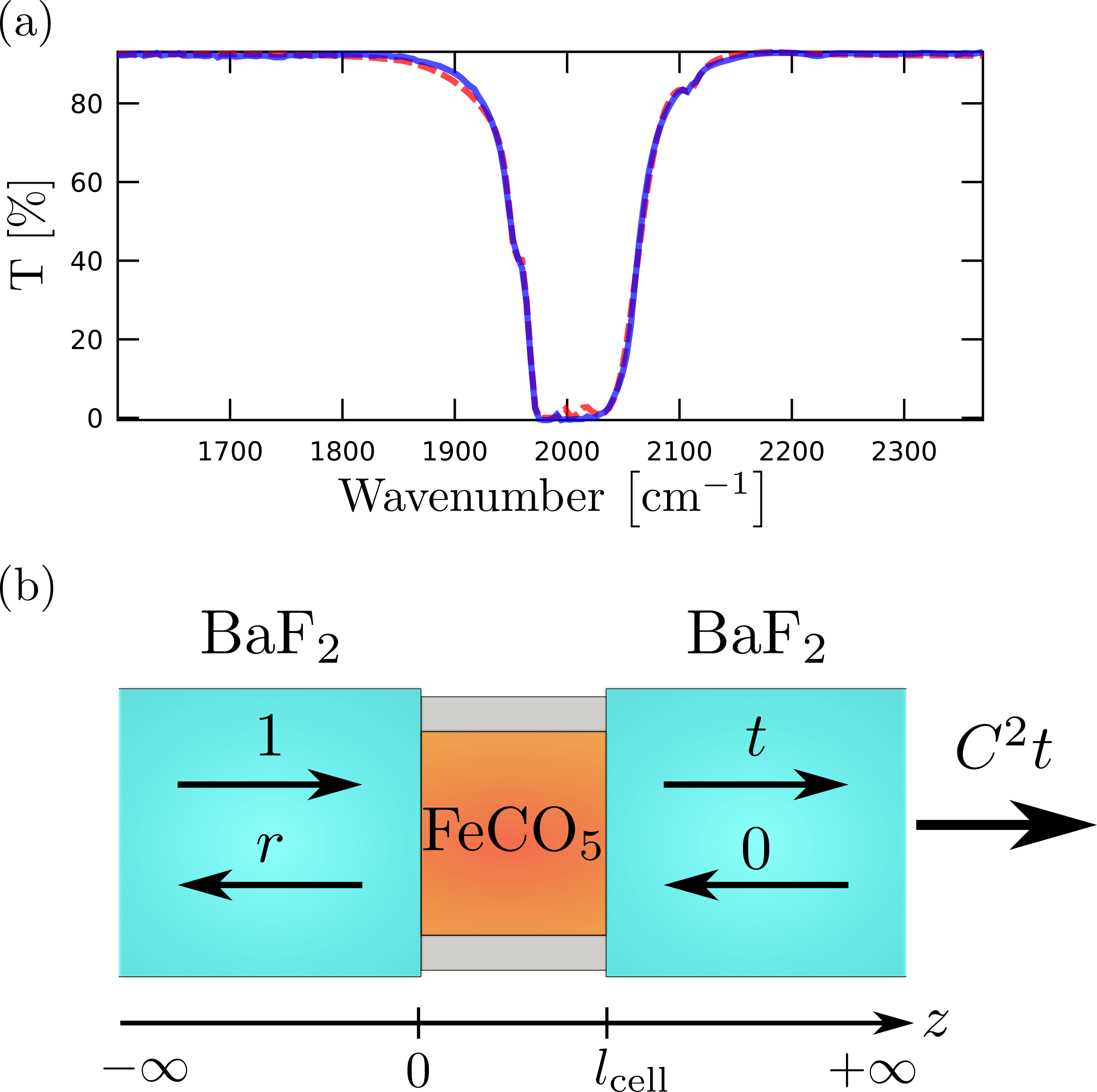}%
\caption{(a) Measured transmission spectrum of the BaF$_2$ flow cell
 filled with FeCO$_5$ (solide blue curve) compared to its transfer
 matrix fit (dashed red curve). (b) Schematic representation of the
 modeled flow cell of thickness $l_{\textrm{cell}}$ for transfer
 matrix calculations. The gray rectangles represent the Mylar spacer
 separating the cell windows. The incident 
 field to the left of the structure has an amplitude of 1. By
 definition, the reflected field has an amplitude $r$ while the field
 propagating to the right of the structure has an amplitude $t$. 
 The transmitted field outside the flow cell has an amplitude $C^2t$, 
 where $C$ is the transmission coefficient of a bare BaF$_2$ window
 measured at the considered wave vector. \label{fig:cell}}
\end{figure}

We first measure the IR transmission spectrum of a dilute Fe(CO)$_5$
solution (10 wt \% in toluene) injected in a barium fluoride BaF$_2$
flow cell. BaF$_2$ windows were preferred to ZnSe for this experiment
 as their lower refractive index minimizes Fabry-P\'erot
 modulations of the transmission spectrum. 
As shown in Fig.~\ref{fig:cell}(a), the spectrum consists of a strong
inhomogeneously broadened absorption peak at $2000\,\textrm{cm}^{-1}$.
To fit this spectrum with the transfer matrix method, we model
the complex refractive index of Fe(CO)$_5$ with a multi-Lorentzian
function:
\begin{equation}
\tilde{n}(k) = \sqrt{n_b^2-\sum_{j=1}^N\mathcal{L}(f_j,k_{0j},\Gamma_j)},
\end{equation}
where $n_b$ is the background refractive index, and
$\mathcal{L}(f_j,k_{0j},\Gamma_j) = f_j/(k^2-k_{0j}^2+ i k\Gamma_j)$,
with $f_j$ the oscillator strength, $k_{oj}$ the resonance wave vector
and $\Gamma_j$ a phenomenological damping constant. It should however be
noted that no specific meaning can be attributed to these
individual Lorentzians since the only criterion here is to reproduce
accurately the flow cell transmission spectrum.
A good fit to this spectrum was obtained using 7
Lorentzians as shown in Fig.~\ref{fig:cell}(a). In this fitting process, the
cell length was also left a free parameter. 
Moreover, the calculations were performed assuming semi-infinite
BaF$_2$ cell windows, their actual
thickness being $\sim\!3\,\textrm{mm}$ (see
Fig.~\ref{fig:cell}(b)). Corrections for the front 
air/BaF$_2$ and rear BaF$_2$/air interfaces where done by using the
bare BaF$_2$ window transmission spectrum.
The resulting fitted parameters are reported in Table~\ref{Table:FitCell}. 

\begin{table*}
\caption{\label{Table:FitCell}Fitted parameters for the transmission
  spectrum of the BaF$_2$ flow cell filled with FeCO$_5$.}
\begin{ruledtabular}
\begin{tabular}{cccccccccccccccccccccccc}
 & $l_{\textrm{cell}}$\footnote{Cavity length in $\mu$m.} & $n_b$ & $f_1$\footnote{All oscillator strengths $f_j$ are in $10^5\,\mu$m$^{-2}$.} & $k_{01}$\footnote{All resonant wave vectors $k_{0j}$ are in $10^3\,\mu$m$^{-1}$.} & $\Gamma_1$\footnote{All phenomenological damping constants $\Gamma_j$ are in $\mu$m$^{-1}$.} & $f_2$ & $k_{02}$ & $\Gamma_2$ & $f_3$ & $k_{03}$ & $\Gamma_3$ & $f_4$ & $k_{04}$ & $\Gamma_4$ & $f_5$ & $k_{05}$ & $\Gamma_5$ & $f_6$ & $k_{06}$ & $\Gamma_6$ & $f_7$ & $k_{07}$ & $\Gamma_7$\\\hline
& $2.00$ & $1.46$ & $3.02$ & $1.98$ & $5.31$ & $0.69$ & $2.02$ & $22.2$ & $0.16$ & $1.95$ & $15.4$ & $0.70$ & $1.99$ & $41.6$ & $0.36$ & $2.00$ & $9.43$ & $0.01$ & $2.10$ & $20.0$ & $4.32$ & $19.8$ & $4.17\cdot10^4$
\end{tabular}
\end{ruledtabular}
\end{table*}

\begin{figure}
\includegraphics[width=\columnwidth]{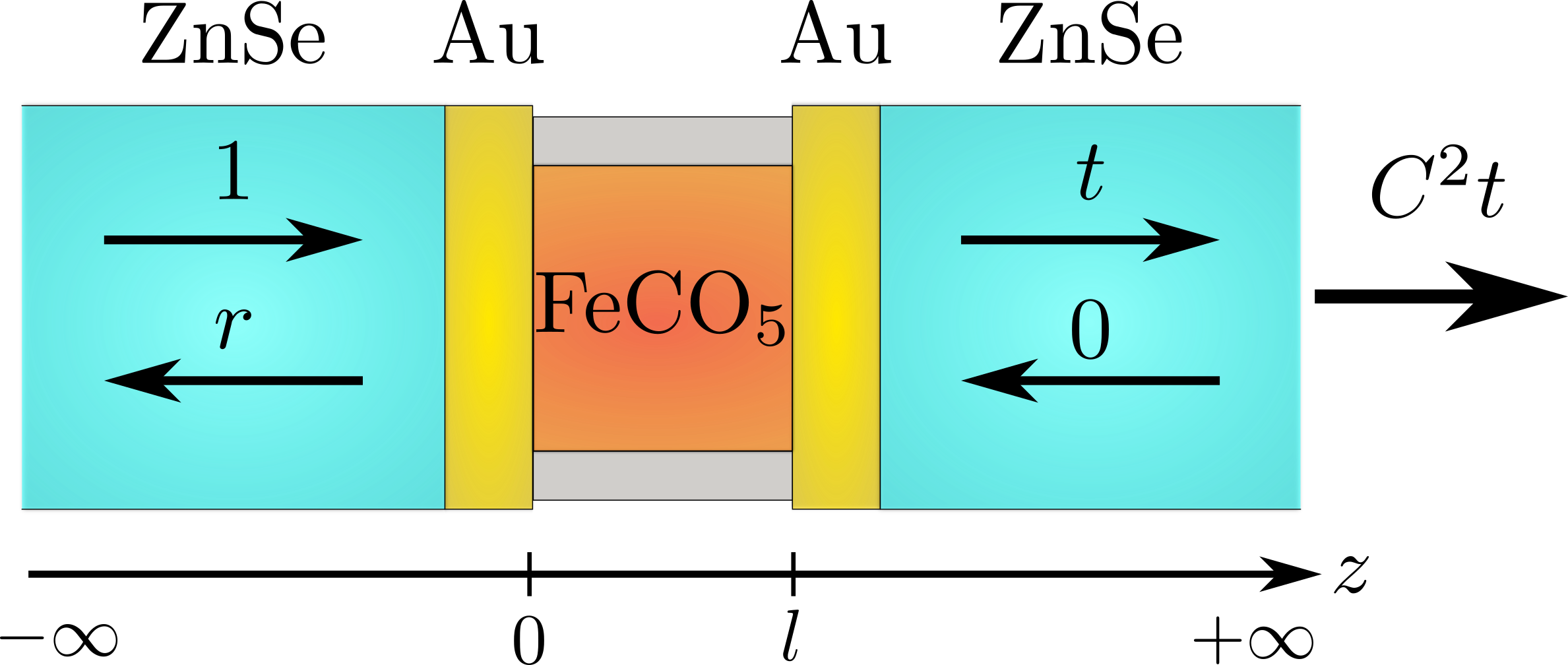}%
\caption{Schematic representation of the
 modeled cavity flow cell of thickness $l$ for transfer
 matrix calculations. The gray rectangles represent the Mylar spacer
 separating the cavity Au mirrors. Those Au mirrors are $13\,\textrm{nm}$
 thick. The field amplitudes are defined as in Fig.~\ref{fig:cell},
 where this time $C$ is the transmission coefficient of a bare
 ZnSe window. \label{fig:cavity}} 
\end{figure}

We now use the fitted refractive index of Fe(CO)$_5$ to model the
transmission spectrum of the ZnSe cavity flow cell, again using the transfer
matrix method. This time the only adjustable parameter is the
exact cavity length $l$, starting from a value of $6\,\mu\textrm{m}$
given by the manufacturer of the Mylar spacer. Indeed, as shown in
Fig.~\ref{fig:cavity}, it is the Mylar spacer thickness that defines
the length of the cavity. 
The best fit value is found to be $l=6.850\,\mu\textrm{m}$.
The cavity Au mirrors are $13\,\textrm{nm}$ thick as fixed by
the sputtering parameters and their refractive indices are taken from
Raki\' c {\it et al.} \cite{rakic} with thickness corrections \cite{ShalabneyNature}. 
As before, the modeling is done assuming semi-infinite
windows, and then correcting for the ZnSe/air interfaces (see Fig.~\ref{fig:cavity}).
The results are shown in  Fig.~2 of the main text.

The free spectral range (FSR) of the fitted cavity transmission
spectrum, as determined by the peak-to-peak frequency spacing in a
non-dispersive spectral region ($5000-7000\,\textrm{cm}^{-1}$), is
$\Delta\nu = 492\textrm{cm}^{-1}$. We note that this FSR is {\textit{not}}
directly related to the cavity length by the usual formula
FSR $=1/2n_bl$ because of the finite skin-depth of the Au cavity
mirrors. Indeed, using this expression and the value of the FSR, we
would expect a cavity thickness $\tilde{l} = 6.932\,\mu\textrm{m}$. The
over-estimation factor with respect to the actual cavity thickness is
$\alpha = \tilde{l}/l = 1.012$. This factor only depends on the
refractive indices of the Au/Fe(CO)$_5$ and Au/ZnSe interfaces and on the Au mirror
thickness. Thus, we will use it to correct the relationship between FSRs
and cavity thicknesses in what follows: $\Delta\nu =1/2\alpha n_bl$.

\section{Vibrational strong coupling of CS$_2$}  \label{AB}

\begin{figure}
\includegraphics[width=\columnwidth]{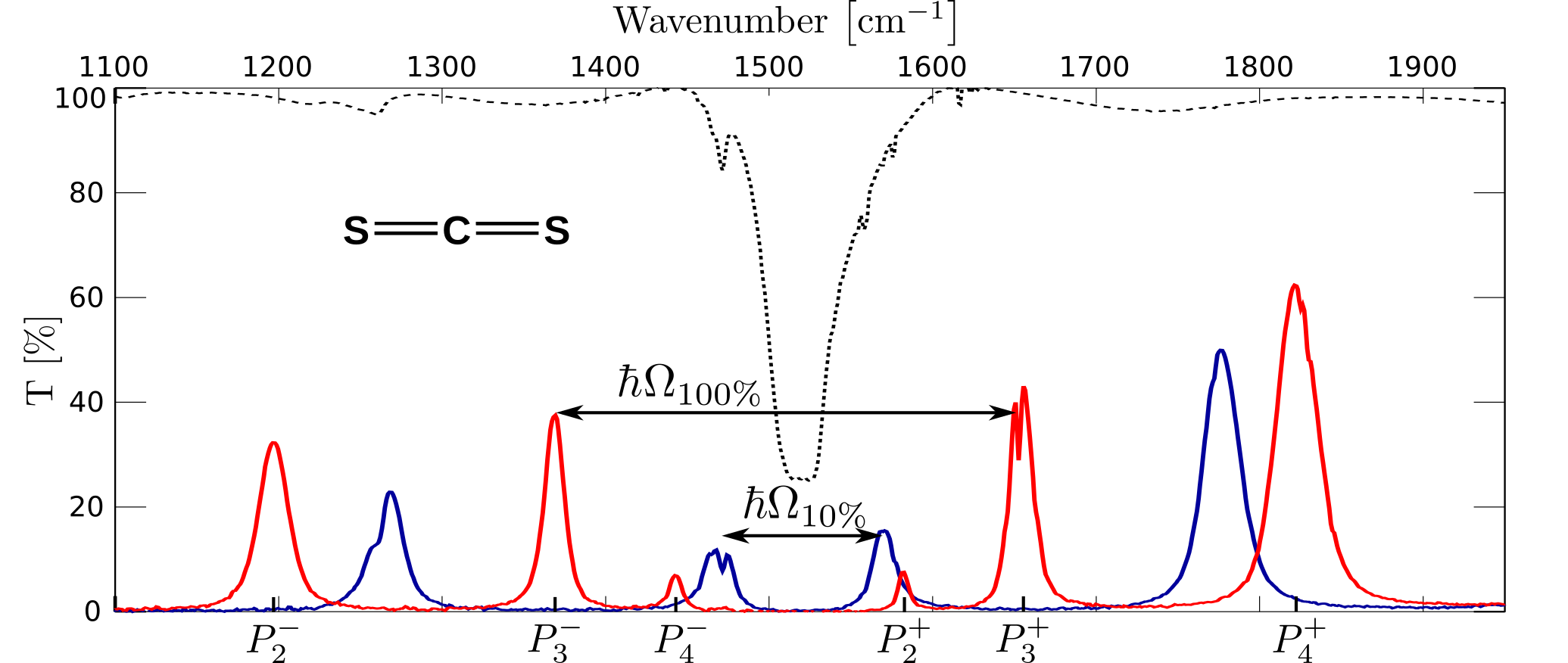}%
\caption{IR spectrum of dilute CS$_2$ (10 wt \% in toluene, black dashed
 line) and corresponding spectrum a resonant cavity (blue
 spectrum). When pure liquid CS$_2$ (100 wt \%) is injected in the cavity,
 multiple splittings are observed with a resonant mode
 splitting $\hbar\Omega_{100\%}\sim\!287\,\textrm{cm}^{-1}$ (red
 curve). The polaritonic modes labeling is the same as in the main
 text. 
 \label{fig:figS3}} 
\end{figure}

Similarly to the case of Fe(CO)$_5$, the spectrum of strongly coupled CS$_2$ goes from the
normal double peaked spectrum at low concentration with a mode
splitting $\hbar\Omega_{10\%}\sim\!100\,\textrm{cm}^{-1}$ to multi-mode
splittings for a pure liquid with
$\hbar\Omega_{100\%}\sim\!287\,\textrm{cm}^{-1}$, as shown in
Fig.~\ref{fig:figS3}. 
This large splitting amounts to ca. 19\% of the vibrational transition
frequency. Here the cavity is tuned to the asymmetric stretching mode of CS$_2$ 
(structure in inset of Fig.~\ref{fig:figS3}) \cite{Plyler}. 

\begin{figure}
\includegraphics[width=0.9\columnwidth]{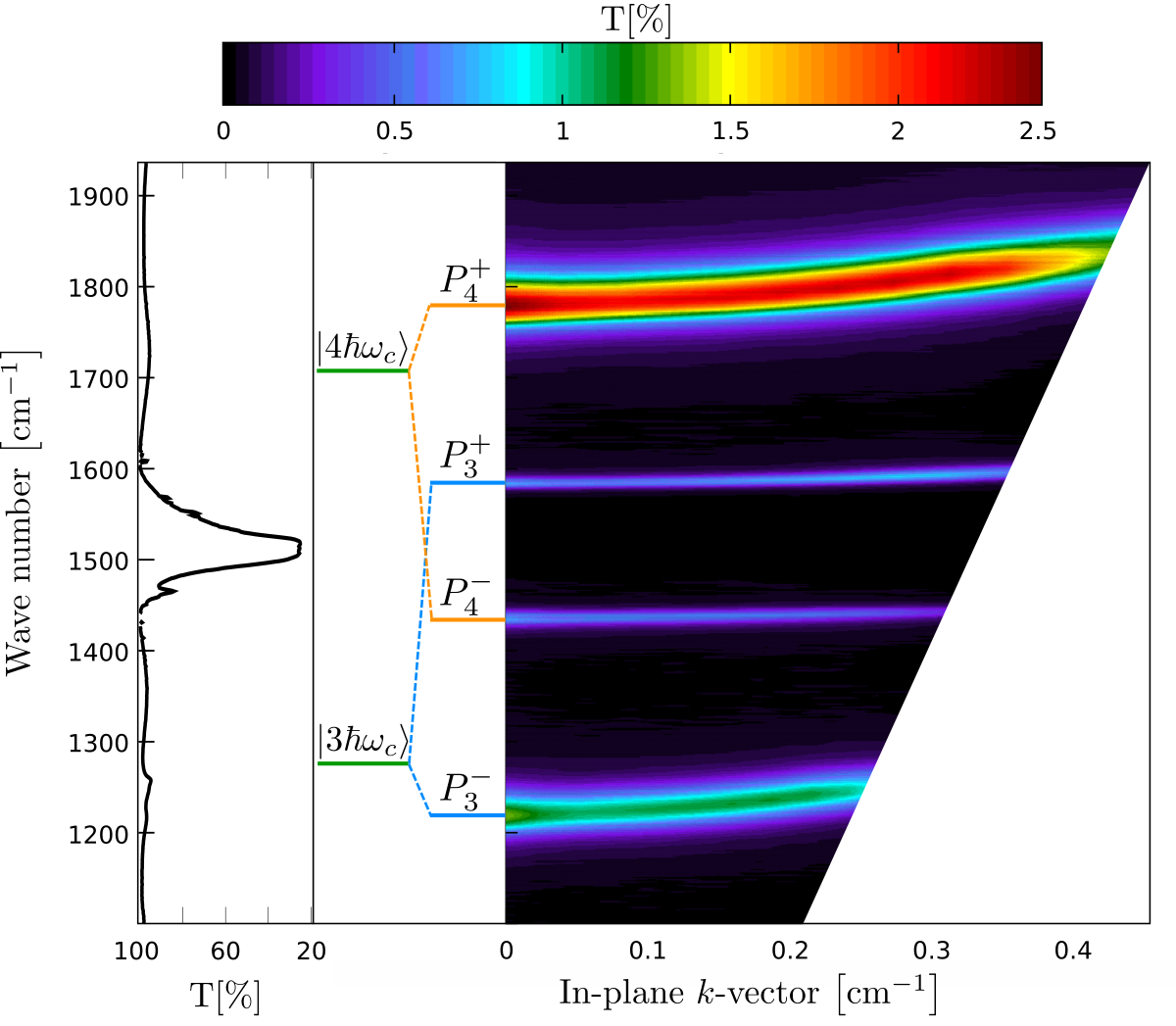}%
\caption{Left panel: dilute CS2 transmission spectrum. Central panel:
  schematic illustration of the off resonant coupling between the CS
  asymmetric stretching mode and the 3rd and $4^{th}$-order optical
  modes of a FP cavity. Right panel: dispersion spectrum measured by
  angle dependent IR transmission spectroscopy (unpolarized,
  $0-22^{o}$) of pure CS2 in the cavity. \label{fig:figS4}}  
\end{figure}

As discussed in the main text, the high absorbance of the molecules can lead to
multiple polaritonic states involving the off-resonance modes of the
cavity. By a similar process, one can reach the extreme case in which
polaritonic states are observed even when no optical mode is resonant
with the vibrational transition. This is illustrated in Fig.~\ref{fig:figS4} where
pure liquid CS$_2$ is injected into the cavity tuned so that the CS
stretching mode energy lies in the cavity free spectral range, between the $3^{rd}$ and
$4^{th}$ modes as shown schematically in the left panel.  
The resulting transmission spectrum displays an off-resonance coupling
with a skewed energy level distribution, with two new dispersive modes
on either sides of the bare CS stretching band (Fig.~\ref{fig:figS4}, right panel). 
Moreover, it can be noted that the new states formed have an energy
splitting at normal incidence of $\sim\!350\,\textrm{cm}^{-1}$
resulting from both the effect of detuning and coupling
interaction. Again the dispersive behavior of these states 
is direct reflection of their photonic versus vibrational
components. These characteristics of the off-resonant hybrid states
are highly analogous to those resulting from orbital mixing in
organometallic complexes in which the metal-ligand bonding molecular
orbitals have more ligand character and anti-bonding molecular
orbitals have more metal character, as explained by Ligand Field
Theory (LFT) \cite{Figgis}.

\section{Collective vibrational coupling: a model for ultra-strongly coupled oscillators}  \label{AC}

We describe our molecular liquid as an ensemble of $N$ individual
ground-state mechanical oscillators to each of which is associated a
localized vibrational dipole ${\bf p}_i$. This yields a effective density of
polarization $\mathcal{P}({\bf r})=\sum_{i=1}^N{\bf p}_i\delta({\bf r})$
which corresponds to a collective dipole ${\bf P}({\bf 0})=\sum_{i=1}^N{\bf p}_i$ localized on ${\bf r}={\bf 0}$. This collective dipole is coupled to the electric displacement ${\bf D}({\bf r})$ of a
single $m^{th}$-order longitudinal cavity mode of volume $V_c$. In
this effective dipolar point of view and neglecting the inhomogeneities of the
cavity mode profile, the Hamiltonian describing the coupled system writes as
\begin{equation}
H =  H_{\rm cav}^m + H_{\rm vib} -\frac{1}{\varepsilon_0}{\bf D}({\bf 0})\cdot{\bf P}({\bf 0})+\frac{1}{2\varepsilon_0 V_\nu}{\bf P}({\bf 0})^2
\label{eq:GM}
\end{equation} 
where $V_\nu$ corresponds to the intra-cavity volume occupied by the
molecules. 

With a background refractive index $n$ inside the cavity of length $L$, the dispersion of the $m^{th}$-order mode writes as 
\begin{equation}
\omega_{c}^{m} = \frac{c}{n}\sqrt{\left(\frac{m\pi}{L}\right)^2+\left | {\bf k}_\parallel \right | ^2}
\end{equation}
and is therefore parametrized by the conserved in-plane component ${\bf k}_\parallel$ of the light wavevector. This leads to define the cavity field Hamiltonian:
\begin{equation}
H_{\rm cav}^m=\frac{1}{4}\hbar\omega_{c}^{m} \left(Q_{c}^2+P_{c}^2\right)
\end{equation}
where the optical position $Q_{c}$ and momentum $P_{c}$ quadratures are introduced, built from the $a(a^\dagger)$ annihilation (creation) photon operators as
\begin{equation}
\left(
\begin{array}{c}
Q_{c} \\
P_{c}
\end{array}
\right)=\left(
\begin{array}{cc}
1  & 1\\
-i & i
\end{array}
\right)
\left(
\begin{array}{c}
a \\
a^\dagger
\end{array}
\right),
\end{equation}
with $[Q_c,P_c]= 2i$ considering that $[a,a^\dagger]=1$.

Putting aside the rotational excitations which are not resolved in our experiment, the Born-Oppenheimer (BO) approximation enables us to separate the electronic and vibrational intra-molecular modes. We can thus consider that the vibrational dipole associated with each of the CO-stretching mode $i$ of one Fe(CO)$_5$ molecule is merely defined from the dependence of the dipole moment $\langle {\bf p}\rangle _e(Q)_i$ on nuclear coordinates $Q$ within the electronic state $e$ considered. The BO approximation also insures that the vibrational dynamics is performed within the same electronic quantum state -in our case, the electronic ground state of Fe(CO)$_5$. Within this approach, each of the $N$ molecular vibrations are treated in the harmonic approximation (see \cite{ShalabneyNature}) and we thus define the collective vibrational Hamiltonian as
\begin{equation}
H_{\rm vib}=\frac{1}{4}\sum_{i=1}^N \hbar\omega_{\nu , i} \left(Q_{\nu, i}^2+P_{\nu, i}^2\right)
\end{equation}
with $\omega_\nu^i$ the vibrational transition associated with a single oscillator and 
\begin{equation}
\left(
\begin{array}{c}
Q_{\nu, i} \\
P_{\nu, i}
\end{array}
\right)=\left(
\begin{array}{cc}
1  & 1\\
-i & i
\end{array}
\right)
\left(
\begin{array}{c}
b_i \\
b_i^\dagger
\end{array}
\right),
\end{equation}
the vibrational position and momentum quadratures related to the $b_i(b_i^\dagger)$ annihilation (creation) operators of the vibrational mode of the $i^{th}$ molecule. The commutators simply write as $[b_i,b_j^\dagger]=\delta_{i,j}$ and $[Q_{\nu , i},P_{\nu , j }]=2i\delta_{i,j}$.

At room temperature, one only retains low vibrational excitation levels so that the vibrational dipole moment is given by a first-order expansion on the nuclear coordinates\begin{equation}
\langle {\bf p}\rangle({Q})_i=\langle {\bf p}_i\rangle_0+\left(\frac{\partial \langle {\bf p}\rangle}{\partial {Q}_i}\right)_0\cdot {Q}_i
\end{equation}
This expansion is taken with respect to the equilibrium nuclear configuration (indicated by the subscript $0$) in the harmonic mean potential of the electronic ground state of the Fe(CO)$_5$ molecule. The first term corresponds to the static dipole moment of the molecule at this equilibrium nuclear position. This static term cancels out due to the $D_{3h}$ point group symmetry of the Fe(CO)$_5$ molecule. 

The nuclear coordinate associated with the harmonic molecular vibration is described in our model by a position quadrature operator
\begin{equation}
\hat{Q}_i=\sqrt{\frac{\hbar}{2\mu_i \omega_{\nu , i}}}Q_{\nu, i}
\end{equation}
where $\mu_i$ is the reduced mass of the vibrational mode and $Q_{{\rm zpf}, i}=\sqrt{\hbar / 2\mu_i \omega_{\nu , i}}$ the zero-point fluctuation amplitude of the molecular oscillator. 

At this stage, we now assume that all vibrational modes are strictly degenerate in energy and mass $\omega_{\nu , i}=\omega_{\nu , j},\mu_i =\mu_j$. This leads us to the definition of the collective dipole operator
\begin{equation}
\hat{\bf P}({\bf 0}) = \left(\frac{\partial \langle {\bf p}\rangle}{\partial {Q}}\right)_0 Q_{\rm zpf} \sum_{i=1}^N Q_{\nu, i}.
\end{equation}
The operator corresponding to the electric displacement of the
$m^{th}$-order mode is given by \cite{CohenBook}
\begin{equation}
\frac{1}{\varepsilon_0}\hat{\bf D}({\bf
  0})=i\sqrt{\frac{\hbar\omega_c^m}{2\varepsilon_0V_c}}\left( a{\bm
    \epsilon}^m -  a^\dagger{\bm\epsilon^\star}^m\right)
\end{equation}
For the sake of the model's simplicity, we will also assume {\it (i)}
that the dipoles are all perfectly aligned with the polarization
${\bm\epsilon}^m$ of
the intracavity field and {\it (ii)} that $V_c = V_\nu = V$. Under these assumptions, the Hamiltonian (\ref{eq:GM}) becomes 
\begin{eqnarray}
H &=&  H_{\rm cav}^m + H_{\rm vib} \nonumber  \\ 
&& - i\hbar\Omega (a-a^\dagger)\sum_{i=1}^{N}Q_{\nu , i}\nonumber  \\ 
&& +\kappa^2\sum_{i=1}^N Q_{\nu , i}\sum_{j=1}^NQ_{\nu , j}
\label{eq:Hamiltonian}
\end{eqnarray}
with
\begin{eqnarray}
\hbar \Omega &=& \left(\frac{\partial \langle {\bf p}\rangle}{\partial {Q}}\right)_0\sqrt{\frac{\hbar\omega_c}{2\varepsilon_0 V}}Q_{{\rm zpf}}  \\ \kappa^2 &=& \frac{1}{2\varepsilon_0 V}\left(\frac{\partial \langle {\bf p}\rangle}{\partial {Q}}\right)_0^2Q_{{\rm zpf}}^2  \nonumber  \\
&=&\frac{\hbar \Omega^2}{\omega_\nu}
\end{eqnarray}
when at resonance $\omega_c^m=\omega_\nu$.

Given the exceptionally high coupling strength provided by the
molecular liquid, the dipolar self-energy term ${\bf P}({\bf 0})^2 /
2\varepsilon_0 V $ in the full Hamiltonian cannot be neglected in our
description. The contribution of this self-energy can safely be
neglected in the standard regime of strong coupling regime, but as it
is known in the context of intersubband polaritonic modes \cite{Ciuti,Todorov}, it must be fully accounted for in the definition of the equation of motion of the vibrational polaritonic states in the ultra-strong coupling regime. This is the central point that we demonstrate in the context of collective vibrational excitations.

To do so, we adapt the original procedure of Hopfield \cite{Hopfield} (see also \cite{Ciuti}) to the case of an ensemble of vibrational modes. This procedure consists in writing down the equation of motion of a polaritonic annihilation operator defined as a normal mode operator $\chi_\pm$ of the system  
\begin{equation}
[ \chi_\pm, H ] = \omega_\pm \chi_\pm
\end{equation}
where $\omega_\pm$ are the energies associated with the upper $+$ and lower $-$ polaritonic states. 

Dealing with an ensemble of vibrational modes coupled to the cavity field, the definition of the normal mode operator 
\begin{equation}
\chi_\pm = w_\pm a + x_\pm B +y_\pm a^\dagger + z_\pm B^\dagger
\end{equation}
involves collective operators defined as
\begin{eqnarray}
B (B^\dagger ) &=& \frac{1}{\sqrt{N}}\sum_{i=1}^N b_i (b^\dagger_i)  \nonumber \\
 B+B^\dagger&=& \frac{1}{\sqrt{N}}\sum_{i=1}^{N}Q_{\nu , i}.
\end{eqnarray}
From $[ b_i, b_j^\dagger ]=\delta_{i,j}$, these collective operators obey canonical commutation relations 
 \begin{eqnarray}
[ B, B^\dagger] = 1.
\end{eqnarray}
This implies that the normal mode operators will have the simple commutation relations
\begin{eqnarray}
&&[ \chi_\pm, \chi_\pm] = [\chi_\pm^\dagger,\chi_\pm^\dagger]= 0 \nonumber  \\
&&[\chi_\pm,\chi_\pm^\dagger] = 1.
\end{eqnarray}
These definitions also lead to the following commutation rules
\begin{eqnarray}
&&[ B, \sum_{i=1}^{N}Q_{\nu , i} ] = \sqrt{N} , \nonumber  \\
 && [ B^\dagger, \sum_{i=1}^{N}Q_{\nu , i} ] =  -\sqrt{N} , \nonumber  \\
&& [ B, \sum_{i=1}^{N}Q_{\nu , i}\cdot \sum_{j=1}^{N}Q_{\nu , j} ] =  2N \left(B + B^\dagger\right) , \nonumber  \\
&& [ B^\dagger, \sum_{i=1}^{N}Q_{\nu , i}\cdot \sum_{j=1}^{N}Q_{\nu , j} ] = -2N \left(B + B^\dagger\right).
\end{eqnarray}
Using these rules, we derive the matrix for the equation of motion (Hopfield matrix)
\begin{equation}
\left(
\begin{array}{cccc}
\hbar\omega_c^m & i\hbar\Omega_{\rm R}& 0 & i\hbar\Omega_{\rm R} \\ 
-i\hbar\Omega_{\rm R} & \hbar\omega_\nu +2\hbar D &  i\hbar\Omega_{\rm R}& 2\hbar D \\ 
0 &  i\hbar\Omega_{\rm R} & -\hbar\omega_c &  i\hbar\Omega_{\rm R} \\ 
i\hbar\Omega_{\rm R} & -2\hbar D  & -i\hbar\Omega_{\rm R} & -\hbar\omega_\nu -2\hbar D
\end{array} 
\right).
\label{hop.full}
\end{equation}
where $\hbar\Omega_{\rm R}= \hbar\Omega\sqrt{N} $ corresponds to the well-known fact that the collective coupling strength is $\sqrt{N}$ time stronger than in the case of a single vibrational mode. The self-energy term is also enhanced by $N$ since we have $\hbar D= N\kappa^2 = \hbar (\Omega_{\rm R}^2/\omega_\nu)$.

The so-called rotating wave approximation (RWA) of the Hamiltonian
(\ref{eq:Hamiltonian}) amounts to neglecting the off-diagonal blocks of
the Hopfield matrix. Interestingly, as pointed out by
Todorov {\it et al.} \cite{Todorov}, polarization self-interaction must also
be neglected in this regime, as it is a
$\mathcal{O}(\Omega_{\rm R}^2/\omega_\nu)$ order term. The resulting
matrix is
\begin{equation}
\left(
\begin{array}{cccc}
\hbar\omega_c & i\hbar\Omega_{\rm R} & 0 & 0 \\ 
-i\hbar\Omega_{\rm R} & \hbar\omega_\nu & 0 & 0 \\ 
0 & 0 & -\hbar\omega_c & i\hbar\Omega_{\rm R}\\ 
0 & 0 & -i\hbar\Omega_{\rm R}& -\hbar\omega_\nu
\end{array} 
\right)_{\rm RWA}.
\label{hop.RWA}
\end{equation}

\section{Fitting the dispersion data}  \label{AD}

As explained in
the main text, we measure experimentally the transmission spectra of cavities of
different thicknesses, all filled with pure Fe(CO)$_5$. The thickness of each of those cavities is directly
accessible from the value of their FSR by making use of the correction
factor $\alpha$ described in section A. We thus end up
with a set of cavity spectra of various thicknesses, each of them
displaying multiple polaritonic resonances.

\begin{figure}
\includegraphics[width=\columnwidth]{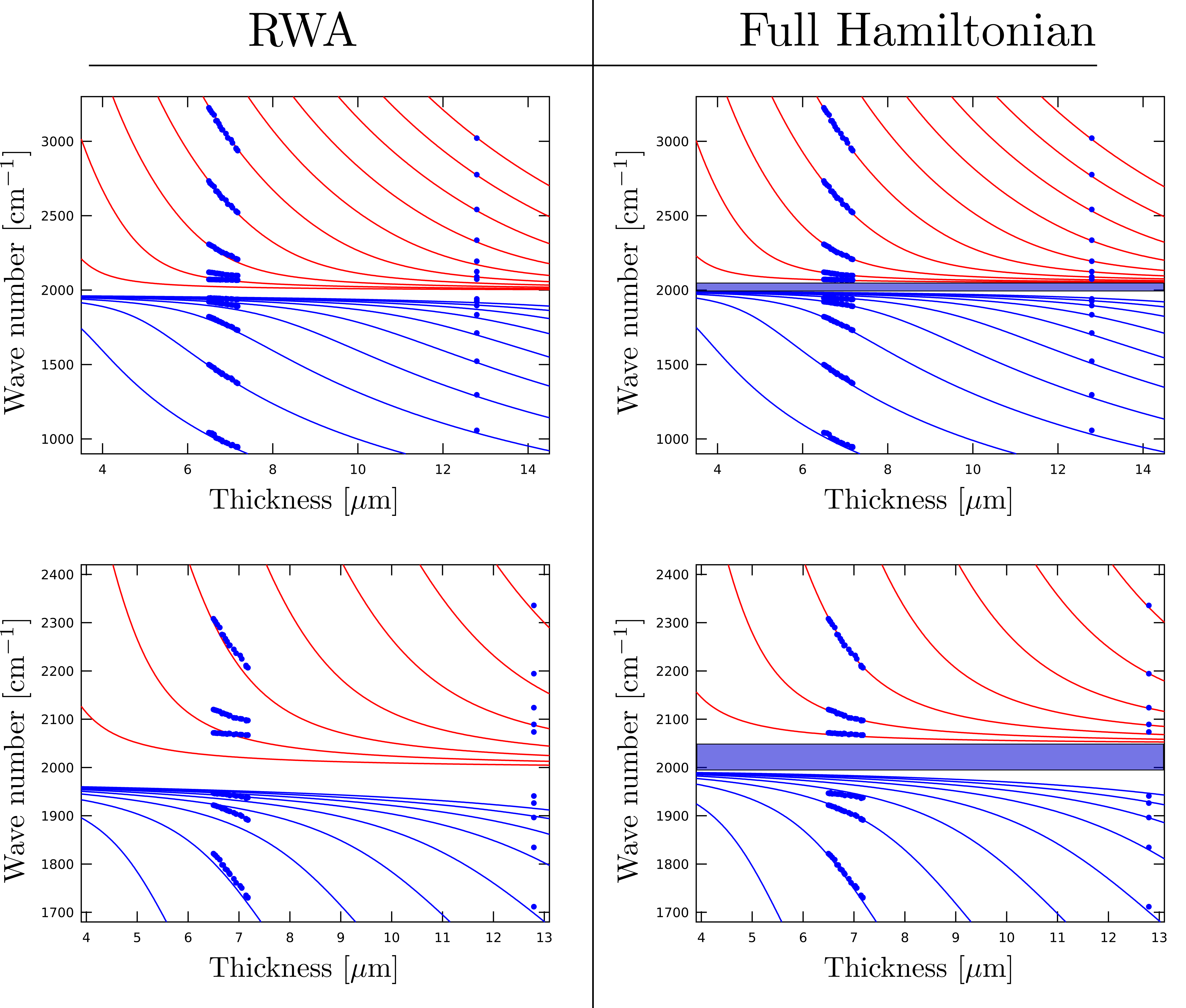}%
\caption{Comparison between the fitting results for the RWA and full
  Hamiltonian models. In both cases, the 10 observable polaritonic
  branches are simultaneously fitted via the Rabi splitting
$\hbar\Omega_{\rm R}$. The bottom row is a blown up on around the
bare molecular vibrational mode energy. \label{fig:figS5}} 
\end{figure}

The resulting multimode polaritonic dispersion curves were compared
successively to the full Hamiltonian model and to the RWA model by
diagonalizing their respective Hopfield matrices (\ref{hop.full}) and (\ref{hop.RWA}). We stress
that all the parameters entering those two models are determined
experimentally, except for the Rabi splitting
$\hbar\Omega_{\rm R}$. Indeed, the 
cavity mode energy $\hbar\omega_c^m$ is directly known from the mode order, the cavity
FSR and the refractive index, while the vibrational energy
$\hbar\omega_\nu$ simply corresponds to the energy of maximal
extinction in the dilute solution transmission 
spectrum of Fig.~\ref{fig:cell}(a). This implies that the Rabi splitting
$\hbar\Omega_{\rm R}$ is the unique free parameter in both
fits. 

In both cases, we search for a minimum of the following quantity
\cite{NLOPT}: 
\begin{equation}
  \chi^2 = \sum_{P_m^\pm}\sum_l (E_{P_m^\pm}(l)-\hat{E}_{P_m^\pm}(l))^2
\end{equation}
where $P_m^\pm$ is the upper (lower) polaritonic branch of $m^{th}$
order, $l$ is the cavity thickness,  
$E$ is the corresponding measured vibrational polaritonic energy and $\hat{E}$ is
the calculated vibrational polaritonic  
energy which depends on the fitting parameter.

The main result of our analysis is the fact that even though both
models match well the data on a broad energy range, the RWA model
is totally unable to reproduce the measured dispersions close to the bare
vibrational mode energy, as shown in
Fig.~\ref{fig:figS5}. As emphasized in the main text, this proves that our system
has genuinely reached the regime of USC.

\bibliography{biblio}

\end{document}